\documentclass[submission,copyright,creativecommons]{eptcs}
\providecommand{\event}{F-IDE Workshop 2014}
\usepackage{breakurl}             

\usepackage{listings}
\usepackage{MnSymbol}
\usepackage{stmaryrd}  
\usepackage{mathpartir}  
\usepackage{color}
\usepackage{times}  

\newcommand{\ocaml}{{\sf OCaml}}
\newcommand{\focal}{{\sf FoCaLiZe}}
\newcommand{\zenon}{{\sf Zenon}}
\newcommand{\zvtov}{{\sf ZvToV}}
\newcommand{\coq}{{\sf Coq}}
\newcommand{\foc}{{\sf Foc}}

\newcommand{\Ocaml}{{\sf OCaml}}
\newcommand{\Focal}{{\sf FoCaLiZe}}
\newcommand{\Zenon}{{\sf Zenon}}
\newcommand{\Coq}{{\sf Coq}}

\lstdefinelanguage{FoCaLiZe}
  {morekeywords={alias, all, and, as, assume, assumed, begin, by, caml,
      collection, coq, coq_require, definition, else, end,
      ex, external, false, function, hypothesis, if, in,
      inherit, internal, implement, is, let, lexicographic,
      local, logical, match, measure, not, notation, of, open,
      on, or, order, proof, prop, property, prove, qed, rec,
      representation, Self, signature, species, step,
      structural, termination, then, theorem, true, type, use,
      with, conclude},
    otherkeywords={->, :, /\\, \\/, =>, \~, \#, ;, \&, \|,=,<>},
    sensitive=false,
    morecomment=[n]{(*}{*)},  
    morecomment=[n]{(**}{*)}, 
    morestring=[b]",
  }

\newcommand{\setlangfocalize}{
\lstset{
  language=FoCaLiZe, tabsize=2, frame=none, breaklines=true,
  basicstyle=\ttfamily, framexleftmargin=1mm, xleftmargin=1mm
}
}
\setlangfocalize

\lstdefinelanguage{MyCoq}
  {morekeywords={Variable, Theorem, Section, assert, End, Qed, Let,
      forall, ex, Chapter, Record, Parameter, Definition, Hypothesis,
      let, match, if, then, else, end, Function, Module, Inductive,
      Require, type, Fixpoint, fix, struct,inversion,discriminate,
      auto, trivial, intro, intros, induction, exact, elim, left,
      right, destruct},
    otherkeywords={->, :, :=, =>, ;, \|,<>,=, /\\, \\/},
    sensitive=false,
    morecomment=[n]{(*}{*)},   
    morecomment=[n]{(**}{*)},  
    morestring=[b]",
  }

\newenvironment{compact-itemize}{
\begin{itemize}
     \setlength{\itemsep}{0.5pt}
     \setlength{\parskip}{0pt}
     \setlength{\parsep}{0pt}}
{\end{itemize}
}

\newcommand{\decldeps}[1] {\lbag #1\rbag}

\newcommand{\defdeps}[1]
  {\lbag\hspace{-0.2cm}\lbag #1\rbag\hspace{-0.2cm}\rbag}

\newcommand{\visuniv}[1]{\mid{#1}\mid}

\newcommand{\mtype}[2]{{\cal T}_{#1}({#2})}

\newcommand{\defdepbytrans}[1]{<^{def}_{#1}}

\newcommand{\mintypenv}{\doublecap}

\newcommand{\eddepsonparam}[3]
        {{\cal D}o{\cal P}_{\textsc{[Expr]}}({#1},{#2})[{#3}]}

\newcommand{\bdepsonparam}[3]
        {{\cal D}o{\cal P}_{\textsc{[Body]}}({#1},{#2})[{#3}]}

\newcommand{\tdepsonparam}[3]
        {{\cal D}o{\cal P}_{\textsc{[Type]}}({#1},{#2})[{#3}]}

\newcommand{\dddepsonparam}[3]{{\cal D}o{\cal P}_{\textsc{[Def]}}({#1},{#2})[{#3}]}

\newcommand{\udepsonparam}[3]
        {{\cal D}o{\cal P}_{\textsc{[Univ]}}({#1},{#2})[{#3}]}

\newcommand{\pdepsonparam}[3]
        {{\cal D}o{\cal P}_{\textsc{[Prm]}}({#1},{#2})[{#3}]}

\newcommand{\mbody}[2]{{\cal B}_{#1}({#2})}

\newcommand{\kw}[1] {\mbox{\bf #1}}

\title{\focal: Inside an F-IDE}
\author{Fran\c{c}ois Pessaux
\institute{ENSTA ParisTech - Palaiseau, France}
\email{francois.pessaux@ensta.fr}\\
\vspace{-0.3cm}
{\scriptsize This work is supported by the project BGLE - DEPARTS}
}
\def\titlerunning{\focal: Inside an F-IDE}
\def\authorrunning{F. Pessaux}
\begin{document}
\maketitle

\vspace{-0.4cm}
\begin{abstract}
For years, {\bf I}ntegrated {\bf D}evelopment {\bf E}nvironments have
demonstrated their usefulness in order to ease the development of software.
High-level security or safety systems require proofs of compliance to
standards, based on analyses such as code review and, increasingly nowadays,
formal proofs of conformance to specifications.
This implies mixing computational and logical aspects all along the
development, which naturally raises the need for a notion of {\em Formal} IDE.
This paper examines the \focal\ environment and explores the implementation
issues
raised by the decision to provide a single language to express specification
properties, source code and machine-checked proofs while allowing incremental development and code reusability.
Such features create strong dependencies between functions, properties
and proofs, and impose an particular compilation scheme, which is described
here. The compilation results are runnable \ocaml\ code and
a checkable \coq\ term. All these points are illustrated through a
running example.
\end{abstract}

\section{Introduction}

Thanks to Wikipedia, an Integrated Development Environment 
 is a ``software application that provides comprehensive facilities to
 computer programmers for software development''. Such environments
 do not provide in general  all the necessary  tools for critical software
 developments as safety, security and privacy standards require the use
 of formal methods all along the development cycle to achieve
 high levels of assurance. Every critical system must be submitted to an
 assessment process before commissioning. It consists in a rigorous
 examination of the demonstration, given by the developer,  that  the implementation complies
with the specification requirements. Specifications, source and object
 codes, automated verifications and mechanically-checked (or not)
 proofs are scrutinized and, if needed, validation is re-done with
 other tools.  This assessment process takes a lot of time and is
 expensive.

The \foc\ environment was created ten years ago ({\cite{BoulmeCalc99}) by
T. Hardin and R. Rioboo as a laboratory language to study ``how to
provide comprehensive facilities''  to developers of critical systems,
complying to high-level rates of standards like \cite {IEC-61508,EN-50128,
CC2}. At this moment, only B, Z and some tools based on algebraic
data types were used in such developments. The idea was to couple the
programming facilities of \ocaml\ with the capabilities of the theorem prover
\coq\ while avoiding the use of some complex features of these
languages. The help provided to developers by object-oriented features
like inheritance and late-binding, by abstraction features like
modules, by parametrisation features like  functors was already widely
recognized in the software development world. Such features were
required in the \foc\ specification to manage not only source code but
also requirements and proofs. Their possible codings in \coq\
were studied\cite{BoulmePhD00} and some possibilities of logical
inconsistencies brought by mixing these features were
identified. Some of them were avoided by using a dedicated concrete
syntax. The remaining ones require a dedicated static analysis to be
eliminated.  This was the first version of the ``dependency
calculus'', studied by V. Prevosto\cite{PrevostoPhD03}, who also designed
the first compiler of the language (which name evolved in \focal).
At that time, the programming language was a
pure functional strongly typed  kernel, 
proofs were done directly in \coq. Then, a
language for proofs was introduced  to enable the use of the automated theorem
prover \zenon\cite{ZenonSite} and automatically translate them into proof
terms of \coq.


\focal\ is the current version of \foc. It provides a logical language
to express requirements as first-order formulas depending on
identifiers present in the context, powerful data-types and
pattern-matching, object-oriented and modular features, a language of
proofs done by Zenon, which may use unfolding of functions defined by
pattern-matching.  From the whole development, the compiler issues a
source file written in \ocaml\ and a proof term submitted to \coq. The
two files are kept close to the \focal\ source to ease traceability
and the compiler guarantees that, if the \focal\ source can be
compiled, then its \ocaml\ translation will be runnable and its \coq\
translation will be checkable. This is a way to avoid returning errors
from target languages, which would be unacceptable for the user.

In this paper, we present the main  features of the compiler, trying to
explain why  they permit to improve the adaptability of \focal\ to the
needs of formal developments. More returns on user experience
using \focal\ and help provided by this environment can been found in
\cite{Ayrault:2009:DGV:1575637.1575640,Ayrault:2009:DLC:1618874.1619017,
PrevostoCalculemus03}.

The first specificity of \focal\ is the
mixing of logical and computational aspects which create a lot of
dependencies between elements of a development.   Definitions are 
considered as terms of the logical language,  properties can embed
names of functions being unfolded in proofs. Late-binding
allows to change definitions thus altering proofs done with the old
version. The second specificity of the current version of \focal\ is
the maximizing of sharing between computational and logical codes, at
the source level, 
through  the use of inheritance, late-binding and modularity which
also adds dependencies. To keep this maximal sharing at object levels, 
 a rather new usage of lambda-lifting relying on a
static analysis of dependencies is used by the compiler and presented
here. 

The first section  presents  the paradigms \focal\ is based on and introduces
a running example to support further discussions. In the next section, the
notion of dependency is introduced, followed by the formal description of
their computation. The last section sketches the code generation model on the
basis of the example.

\section{From the \focal\ Point of View}
\label{ref-design-and-motivs}


\subsection{Semantical Framework}
Specification requirements and implementations should be together
expressible in any FIDE, the first one by logical language, the second
one by a programming language. If these two languages can be related
through a single semantical framework, then the demonstration of the
conformance of developments to requirements can be
facilitated. Choosing an imperative style related to a Hoare logical
language leads to some difficulties when composing specifications and
unfolding function bodies, due to the lack of referential
transparency. So \focal\ is
built upon a pure functional language. Then, whatever is the logic, functions 
can be consistently unfolded in proofs and thus, can be used
without restriction to express requirements (called here {\em properties}) and proofs.

Side-effects are however possible but they have to be confined into dedicated modules, separated
from the rest of the development.   Properties of
functions making side-effects can be separately demonstrated and
rendered as contracts.

Static strong typing, rich data-types and
pattern-matching ease source coding and error detection. This is why
the \focal\ programming (sub-)language is very close to a functional
kernel of the \Ocaml\ language. The logical
sub-language  offers first-order quantifiers and the ability to use
identifiers present in the context in formulas. So formulas can
contain free variables, which  however are bound in
the development context. These formulas are indeed formulas of a
dependent type theory and are translated as such in \Coq, relying on the
Curry-Howard isomorphism : data types are translated onto types,
properties onto types and 
functions on terms.  
 Having our
own logical syntax instead of the one of a dependent theory like \Coq\
allows (not only) to restrict formation rules of terms. For example, properties
can use function names but functions cannot use property names. 


\subsection{Incremental Specification and Development through a Running Example}
\label{ref-example}
To support further discussions, this section will gradually introduce
a running example. It represents a ``monitor'' which inputs a value and
outputs a validity flag indicating the position of the value with
regards to two thresholds, and this value or a default one in case
of problem. A fleshed-out version of this example was
used to develop a generic voter used in critical
software \cite{Ayrault:2009:DGV:1575637.1575640}.

\medskip
\label{ref-increm-spec-n-dev}
Object-oriented features are often offered by IDE as they have the
reputation to help software development. Here, assessment also should
be helped and some of these features like complex rules of visibility (tags private,
friend, etc.) and possibilities of runtime late-binding  can weight
down this process. \focal\ proposes two object-oriented features:
inheritance and late-binding. 
We call here {\em inheritance} the
ability of introducing, at any stage of the development,
properties, functions and proofs (called {\em methods} in \focal) that
these new items contribute to
fulfill previously stated requirements. As usual, {\em late-binding} is the
possibility of introducing the name, and here the type, of
a function while deferring its definition. This allows to adapt
definitions along the inheritance tree and enhances
reusability.

\noindent
\begin{minipage}{6cm}
{\tiny
\begin{lstlisting}
species Data =
  let id = "default" ;
  signature fromInt : int -> Self ;
end ;;
\end{lstlisting}
}
\end{minipage}\hskip1cm
\begin{minipage}{8.5cm}
{\tiny
\begin{lstlisting}
species OrdData =
  inherit Data ;
  signature lt : Self -> Self -> bool ;
  signature eq : Self -> Self -> bool ;
  let gt (x, y) = ~~ (lt (x, y)) && ~~ (eq (x, y)) ;
  property ltNotGt : all x y : Self, lt (x, y) -> ~ gt (x, y) ;
end ;;
\end{lstlisting}
}
\end{minipage}

The component ({\em species}) {\tt Data} simply says that the input is coded
by an integer, converted by {\tt from\_int} to a value of the
internal representation, which is denoted by {\tt Self}. The species
{\tt OrdData} inherits from {\tt Data}, it declares two functions ({\tt lt}
and {\tt eq}), defines a derived function {\tt gt} and states a
property {\tt ltNotGt} using both declared and defined methods.

As function types are parts of
specifications, late-binding cannot change them. But late-binding can
invalidate proofs done with an ``old'' definition and the compiler
must manage this point. Late-binding also allows stating properties and delay
their proofs, while the former are already usable to write other proofs.
Restricting the access to some elements is needed and
handled at the {\em component} level.

 A \focal\  component 
collects not only types, declarations and definitions but
also related properties and proofs. 
 Inside a component, the manipulated
data-types have a double role: a programming one used in type
verification, a logical one when translated to \Coq\ (this double view
has to be managed by the compiler and it is not always straightforward
as seen further). 
 To simplify the \focal\
model, all data-types introduced in  a component are grouped  (by the way of
product and union types) into a single  data-type called
{\em representation}). It gives to this notion of component a flavor of
Algebraic Data Type, a notion which has proven its usefulness in
several IDE (e.g. \cite{MaudeSite}). The representation can be just a type variable (i.e. be not yet given an effective implementation),
whose name serves in declarations. It  can be instantiated by
inheritance (but not redefined) to allow typing of definitions. 

{\tiny
\begin{lstlisting}
species TheInt =
  inherit OrdData;
  representation = int ;
  let id = "native int" ;
  let fromInt (x) : Self = x ;
  let lt (x, y) = x <0x y ;
  let eq (x, y) = x =0x y ;
  proof of ltNotGt = by definition of gt property int_ltNotGt ;
end ;;
\end{lstlisting}
}

The species {\tt TheInt} defines the
representation ({\tt int}) and functions
already declared, then proves the property by unfolding {\tt gt} and using
a property found in the file {\tt basics} of the standard library (this
proof, which could be done in {\tt OrdData}, will be used
later).

Functors allow parametrisation by components. Here a component may not
only use functions provided by other components, but also their
properties and theorems.  Parametrisation by values of these parameter
components is of course needed. The link between a parameterised
component and its parameters will be reflected in \Coq\ via the notion
of \Coq\ dependencies.

{\tiny
\begin{lstlisting}
type statut_t = | In_range | Too_low | Too_high ;;

species IsIn (V is OrdData, minv in V, maxv in V) =
  representation = (V * statut_t) ;
  let getValue (x : Self) = fst (x) ;
  let getStatus (x : Self) = snd (x) ;

  let filter (x) : Self =
    if V!lt (x, minv) then (minv, Too_low)
    else
      if V!gt (x, maxv) then (maxv, Too_high)
      else (x, In_range) ;

  theorem lowMin : all x : V, getStatus (filter (x)) = Too_low -> ~ V!gt(x, minv)
  proof =
   <1>1 assume x : V,
        hypothesis H: snd (filter (x)) = Too_low,
        prove ~ V!gt (x, minv)
        <2>1 prove V!lt (x, minv) by definition of filter type statut_t hypothesis H
        <2>2 qed by step <2>1 property V!ltNotGt
   <1>2 qed by step <1>1 definition of getStatus ;
end ;;
\end{lstlisting}
}

The species {\tt IsIn} has a collection parameter {\tt V} and two
value parameters {\tt minv} and {\tt maxv} from {\tt V}. It mainly shows
a proof decomposition into several steps (see \ref{props-n-proofs}),
including a step of induction (here a simple case split) on the union
type {\tt statut\_t}.

 Should the definition of the representation be exposed or
 encapsulated by  the modularity mechanism?  Inheritance and
 late-binding require its exposure, as total encapsulation can
 cumbersome the development task.  On the contrary parametrisation
 asks for its abstraction.  Indeed, a component seeing the data
 representation of its parameters can manipulate this data without
 using the provided functions, hence breaking invariants and
 structural assumptions made by parameters and invalidating the
 theorems they provide.  Abstract definitions of types is not
 sufficient as  properties can still reveal the exact definition of a
 type (a bad point when security properties are considered).  Thus
 \focal\ has two notions of components. {\em Species} which expose their
 representation are only used along inheritance and late-binding
 during design and refinement.
{\em Collections} which encapsulate the representation are
used as species parameters during the integration process.

To avoid link-time errors, any call of an effective species parameter
must be ensured that all functions exported by this parameter are
really defined. To preserve consistency, all exported properties
must have already received proofs. Thus collections can only be created
 by encapsulation of species, called {\em complete species},
in which all declarations have
received a definition and all properties have a  proof. Encapsulation
builds an {\em interface} exposing only
the name of the representation,  declarations of (visible) functions and
(visible) properties of this species. The compiler guarantees that all
exposed definitions and theorems have been checked and that the
interface is the only way to access collection items.

{\tiny
\begin{lstlisting}
collection IntC = implement TheInt ; end ;;
collection In_5_10 = implement IsIn (IntC, IntC!fromInt (5), IntC!fromInt (10)) ; end ;;
collection In_1_8 = implement IsIn (IntC, IntC!fromInt (1), IntC!fromInt (8)) ; end ;;
\end{lstlisting}
}

The species {\tt TheInt} being {\em complete}, it is submitted to
encapsulation ({\tt implement}) to create the collection {\tt IntC}.
This latter is then candidate to be effective argument of {\tt IsIn}'s
parameter {\tt V} and to apply its method {\tt fromInt} to
provide effective values for the {\tt minv} and {\tt max} parameters.
Hence it can used to create other collections, {\tt In\_5\_10} and
{\tt In\_1\_8}.}

From a developer's point of view, species serve to make an incremental
development. Collections, mostly used as effective parameters of
species, are used to assemble separate parts of the development. As
writing species also needs some primitive notions like booleans,
integers, \focal\ has a standard library which provides these
low-level bricks, possibly along with proved properties but without
encapsulation. The absence of encapsulation is wanted in such a case
and allows manipulating basic data-structures as native constructs
while however having properties. As long as the considered datatypes
(i.e. type definitions here) do not have any invariant to preserve,
there is no risk of inconsistency by revealing their effective structure.
The standard
library consists in source files that can be accessed by an opening
mechanism and is not concerned by the species/collections
mechanism. From the development and assessment points of view, files
of the standard library are just on-the-shelf components granted by
the system (or their provider).

\subsection{Properties and Proofs}
\label{props-n-proofs}

A proof, intended to be mechanically checked, must be decomposed into
small enough steps to be checkable by the prover.  Our first try (\foc)
to directly use the script language of \Coq, revealed several
drawbacks.  First, it required the user to have a deep knowledge of
\Coq\ to powerfully use it and to understand error messages. Second,
the user had to be aware of the compilation of \focal\ elements to
\Coq. Third, the proofs too deeply depended on \Coq.  In \focal\ an
intermediate step to do proofs has been  introduced. It
is based on natural deduction which, being reminiscent of
mathematical reasoning, is  accessible to a non-specialist without too much
effort. It uses the automated theorem prover \Zenon. A proof is  a hierarchical decomposition into
intermediate steps\cite{LamportProof} unfolding definitions,
introducing subgoals and assumptions in the context until reaching a
leaf, that is, a subgoal which can be automatically handled by \Zenon. When all the leaves have received proofs, the compiler
translates the whole proof  to \Coq\  and  builds
the context needed for checking this proof. The following example
shows this decomposition into a  list of steps, always ended by a {\tt qed} step, whose
goal is the parent goal.

{\tiny
\begin{lstlisting}[frame=none]
theorem t : all a b c : bool, a -> (a -> b) -> (b -> c) -> c
proof =
  <1>1 assume a b c : bool,
       hypothesis h1: a, hypothesis h2: a -> b, hypothesis h3: b -> c,
       prove c
    <2>1 prove b by hypothesis h1, h2
    <2>2 qed by step <2>1 hypothesis h3
  <1>2 qed by step <1>1
\end{lstlisting}} 

The proof has two outer steps {\tt <1>1} and {\tt <1>2}.
Step {\tt <1>1} introduces hypotheses {\tt h1}, {\tt h2},
{\tt h3} and the subgoal {\tt c}. It is proved by a 2-step subproof.
Step {\tt <2>1} uses {\tt h1} and {\tt h2} to prove {\tt b}. Step {\tt <2>2}
uses {\tt <2>1} and {\tt h3} in order to prove {\tt c}.
Step {\tt <1>2} ends the whole proof.

Proofs done in a given species are shared by all species inheriting
this one. As late-binding allows redefinition of functions, proofs
using the ``old'' definition are no longer correct, must be
detected as soon as the redefinition is compiled, reverted to the
property status and should be done again.  This link between proofs
and definitions is a particular case of  {\em dependencies} between
elements of the user development. The section \ref{deps-user-code} intuitively
introduces this concept while the section \ref{deps-comput} formally describes
the related calculus. 

Specification requirements such as the safety/security ones can be
split into finer ones (see
\cite{Ayrault:2009:DLC:1618874.1619017}) and proved under the
hypotheses that these finer properties
hold\cite{PrevostoCalculemus03}. Thus, \Focal\ allows to do
proofs {\em just in time} using not already proved properties and 
 guarantees  that they will be proved later in
a derived complete species. This can help early detection of specification
errors. Moreover, some properties can be granted by other means
(contracts, verification tools, etc.). They can be considered as
theorems by giving them a proof reduced to the keyword {\tt
  admitted}. This is a  needed but dangerous feature as admitted properties can
lead to logical inconsistencies. The use of this keyword is recorded in
the automatically done documentation file provided by the compilation
process. The assessment process must ultimately  check that any occurrence of this
keyword is harmless.

\subsection{The Compilation Process}
In a nutshell,  the compilation process guarantees that the
whole development leads to a runnable code satisfying the
requirements. Its {\em code generation}  translates user code to
a source code  (called here simply
{\em computational code}) in \Ocaml\ and to a term 
(called here {\em logical code}) of \Coq.   Data types and properties
are translated  to \Coq\  types  while definitions and proofs
are translated as terms.  Computational code only contains
declarations and definitions since they are the only material contributing
to an executable.

The translations of the user code are type-checked by \Ocaml\ and
\Coq.  But these checkings arrive too late and
error diagnostics  may be difficult to convey to the user. Thus the
compiler has  its own typing mechanism, to early detect typing
errors and emit comprehensive diagnostics. 

First preliminary tries were done to translate inheritance and
late-binding using oriented-object
features of \Ocaml (\cite{BoulmeCalc99}). But, as \Coq\ does not
propose these features, a different way of compilation was needed to
produce logical code (see \cite{PrevostoPhD03}). These two different
compilation schemes
jeopardize traceability as the object codes differ a lot. The current
compiler uses a single method of compilation to both target languages,
inspired by
Prevosto's work and presented in the next section. It  resolves  inheritance and
late-binding  before code generation while  controlling impacts of
redefinitions onto proofs.

\subsection{ \zenon}
The automatic theorem prover \zenon\ is based on a sequent calculus
for first-order logic and has {\sf Fo\-Ca\-Li\-Ze}-specific features such as unfolding of
definitions and inductive types.  Boolean values and logical propositions
are strictly distinct notions in \coq\ but this complexity
is hidden to the \focal\ user: explicit conversions (done by
{\tt Is\_true}) between
{\tt bool} and {\tt Prop} are inserted in the formula sent to 
\zenon. They could be axiomatized but this would blow up the
search-space; instead, \zenon\ has inference rules dedicated to
these conversions.

A proof written in \focal\ is compiled into a tree of \coq\ {\em
  sections}, which translates the natural deduction style into a
natural deduction proof in \coq. Each section first introduces
the hypotheses of its corresponding step, then includes the
  sections of its substeps and ends with the proof of its goal, which is
generated by \zenon\ from the facts of the {\tt qed} step.  Each leaf proof is just temporarily
compiled into a ``hole'' later filled with a \zenon\ proof by
invocation of \zvtov. This tool translates each leaf of the user's proof
 into a \zenon\ input file that contains the goal and
assumptions and definitions listed by the user. Then
\zenon\  outputs  a \coq\ proof which fits right in the
corresponding \coq\ section because every assumption it was given is
available at this point in the \coq\ file.  Once all ``holes'' are
filled, the whole \coq\ source file is sent to \coq\ for verification.

 The only acceptable errors from \zenon\ are simple:
{\em ``out of memory''}, {\em ``time out''}, {\em ``no proof found''}. They
mean that \zenon\ didn't find any proof, either because some hypotheses were
missing (in this case the user may add lemmas to split the proof) or because
the proposition is false. If a proof is found, \coq\  must raise no error when
checking it otherwise there is a bug in \zenon\ or \focal.


\section{Toward Effective Code}
This section begins by an informal presentation of the notion of
dependencies since they strongly impact the form of the target codes
generated by the \focal\ compiler. Next comes the formal computation of
these dependencies. Finally, the code generation model is depicted on the
basis of the above example.

\subsection{Dependencies in User Code}
\label{deps-user-code}
A method {\em depending} on the definition of another method {\tt m}
is said to have a {\em def-de\-pen\-den\-cy} on {\tt m}. Some
def-dependencies are eliminated by a good choice of syntax. For
example, function bodies cannot contain property names nor keywords
for proofs, thus a function cannot def-depend on a proof. There are
only two possibilities of def-dependencies.  First, proofs with a
{\tt by definition of m} step (which unfolds the definition of {\tt m})
def-depend on {\tt m}.  If {\tt m} is redefined, these proofs must be
invalidated. Second, functions and proofs can def-depend on
the representation. Properties must not def-depend on the
representation, as explained by the following example.   The
complete species {\tt Wrong} may be encapsulated into a collection,
whose interface contains the statement {\tt theo}. This one
should  be well-typed in the logical code.  But typing
\lstinline"x + 1" in {\tt theo} requires to identify (unify)
\lstinline"Self" with \lstinline"int". The encapsulation of the
representation prevents it. Thus the species {\tt Wrong} must be
rejected by the compiler.

\noindent
\begin{minipage}{8.5cm}
{\tiny
\begin{lstlisting}[frame=none]
species Wrong =
  representation = int ;
  let incr (x) : Self = x + 1 ;
  theorem theo : all x : Self, incr (x) = x + 1 ;
end ;;
collection Bad = implement Wrong ;;
\end{lstlisting}
}
\end{minipage}\hskip1cm
\begin{minipage}{7cm}
{\tiny {\tt Bad} \ldots bad interface
\begin{lstlisting}[frame=none]
representation : self
incr (x) : Self -> int
theorem theo :
  all x : Self, incr (x) = x + 1 ;
\end{lstlisting}
}
\end{minipage}



Note that function calls do not create def-dependencies and that
encapsulation of collections prevents any def-dependency on methods of
collection parameters.  Thus analysis of def-dependencies must
ensure that proofs remain consistent despite redefinitions and that
properties have no def-dependencies on the representation (in other words,
interfaces of collections should not reveal encapsulated information).

Apart from the def-dependencies of proofs on definitions and of properties
on representations, there are other dependencies called
{\em decl-dependencies}. Roughly speaking, a method {\tt m1} decl-depends
on the method {\tt m2} if {\tt m1} depends on the declaration of
{\tt m2}. The following
example gives a first motivation for their analysis.

\noindent
\begin{minipage}{7.5cm}
{\tiny
\begin{lstlisting}[frame=none]
species S =
  signature odd : int -> bool ;
  let even (n) =
    if n = 0 then true else odd (n - 1) ;
end ;;
\end{lstlisting}
}
\end{minipage}\hskip0.5cm
\begin{minipage}{7.5cm}
{\tiny
\begin{lstlisting}[frame=none]
species T =
  inherit S ;
  let odd (n) =
    if n = 0 then false else even (n - 1) ;
end ;;
\end{lstlisting}
}
\end{minipage}

In {\tt S}, {\tt even}  is at once declared and defined, so
 its type can be inferred by the type-checker, using 
 the type of {\tt odd}. Thus {\tt even} decl-depends on
{\tt odd} but {\tt odd} does not depend on  {\tt even}. 
In {\tt T}, defining {\tt odd}  creates a
decl-dependency of {\tt odd} on {\tt even} and an implicit recursion
between them.  To keep logical consistency, such an implicit recursion must be
rejected. Recursion between entities must be declared (keyword {\tt
  rec}). The compiler has to detect any cycle in dependencies through the
inheritance hierarchy.

More generally, a function {\tt m} decl-depends on {\tt p} if
{\tt m} calls {\tt p}, a property {\tt m} decl-depends on {\tt p} if
typing of {\tt m} in the logical theory requires {\tt p}'s type, a proof
decl-depends on {\tt p} if it contains a step {\tt by property p} or
an expression whose typing needs {\tt p} and, recursively, {\tt m}
decl-depends on any method upon which {\tt p} decl-depends and so
on. Def-dependencies are also decl-dependencies. These cases are not
the only cases of decl-dependencies. The first version of this
calculus is described in \cite{PrevostoJAR02}.

\subsection{Dependencies Computation}
\label{deps-comput}
To support generation of well-formed code, the notion of dependencies
must be reinforced and formally studied. 
 The theorem
{\tt ltNotGt} syntactically decl-depends on {\tt gt}, {\tt lt},
{\tt rep} and def-depends on {\tt gt}. Thus, its proof is ultimately compiled to
a \coq\ term, where {\tt gt} is unfolded, making arising the
identifier {\tt eq}. The type of {\tt eq} is needed to \coq-typecheck
{\tt gt} and must be provided through the $\lambda$-lift  {\tt
  abst\_eq} of {\tt eq}. Only $\lambda$-lifting  syntactic
def- and decl- dependencies  would lead to a generated code looking like:


{\tiny
\begin{lstlisting}[language=MyCoq,mathescape=true,frame=none]
Theorem ltNotGt (abst_T : Set) (abst_lt := lt) (abst_gt := OrdData.gt abst_T abst_eq abst_lt) :
  forall x  y : abst_T, Is_true ((abst_lt x y)) -> ~Is_true ((abst_gt x y)).
apply $"Large\ Coq\ term\ generated\ by\ Zenon"$.
\end{lstlisting}}

\noindent where the \lstinline":=" construct binds def-dependencies, and where 
{\tt abst\_eq} is unbound ! Moreover, raising {\tt eq} also exhibits a
def-dependency on the carrier through the one of {\tt eq}.
Dependecies over collection parameters methods suffer from the same
incompleteness. Hence, a process of ``completion'' of syntactically
found dependencies has to be applied before $\lambda$-lifting.


 It
requires to compute the {\em visible universe} of a method {\tt m}
which is the set of methods of {\tt Self} needed to analyze {\tt m}.
Then, the minimal (\coq) typing environment of {\tt m} is built by
deciding, for each method {\tt p} of its visible universe if only its type
must be kept (issueing  $\lambda$-lift of {\tt p}) or if its body is
also needed (hence ``\lstinline":="-binding'' of {\tt p}). Finding the minimal set is
especially important since this allows to reduce the amount of
$\lambda$-liftings of method and collection generators.
A last
completion of the set of dependencies for parameter methods achieves
the building.


\medskip
In the following, we assume that inheritance has been processed, leading
to a {\em normal form} of a species in which all its methods are present,
in their most recent version and well-typed. Although this process is not
trivial (\cite{PrevostoJAR02,PrevostoTPHOL02}), it is out of the scope of
this paper.


\subsubsection{On Methods of {\tt Self}}

Let $S$ be a species. We denote by $\decldeps{x}_S$ (resp. $\defdeps{x}_S$) the
set of methods of $S$ on which the method $x$ decl-depends (resp. def-depends).
This set is obtained by walking the Abstract Syntax Tree looking for
apparition of methods names in expressions (resp. of
\lstinline"by definition" in proofs). Only dependencies on the carrier have to
be checked differently, by typechecking.
As proofs and properties names are syntactically forbidden inside
property expressions and function bodies,  typechecking of properties
and functions requires
only the types of the function names appearing in them  (i.e. appearing in $\decldeps{}$).

Considering theorems proofs (i.e. bodies), def-dependencies can arise
forcing the need to keep some definitions (not only types) to be able to
typecheck. Then, one also needs to make sure that these
definitions can themselves be typechecked.
Moreover, proofs may involve decl-dependencies on logical methods, whose
types are logical statements (i.e. expressions). Methods appearing in such
types must also be typechecked.
\medskip
For all these reasons, the context needs to keep trace of the methods belonging to
the transitive closure of the def-dependency relation, plus the methods on
which these latter decl-depend. This context is called the
{\em visible universe} of a method $x$ and is noted $\visuniv{x}$. In the same
spirit than in \cite{PrevostoPhD03}, $\visuniv{x}$ is defined as follows:

\vspace{-0.2cm}
{\scriptsize
$$
\inferrule
  {y \in {\decldeps x}_S}
  {y \in \visuniv{x}}
\\
\hspace{0.6cm}
\inferrule
  {y \defdepbytrans{S} x}
  {y \in \visuniv{x}}
\\
\hspace{0.6cm}
\inferrule
  { z \defdepbytrans{S} x \\ y \in {\decldeps z}_S}
  { y \in \visuniv{x}}
\\
\hspace{0.6cm}
\inferrule
  { z \in \visuniv{x} \\
   y \in {\decldeps {\mtype S z}}_S}
  { y \in \visuniv{x}}
$$
}

\vspace{-0.5cm}
\noindent where $x \defdepbytrans{S} y$ stands for $y$ def-depends on $x$ by
transitivity and ${\mtype S x}$ stands for the type of $x$ in the
species $S$.

\medskip
From the notion of visible universe, it is possible to define, for a method $x$
of a species, what are the other methods needed to have $x$ well-typed.
\begin{compact-itemize}
\item Methods not present in the visible universe are not required.
\item Methods present in the visible universe on which $x$ doesn't
  def-depend are required but only their type is needed.
\item Methods present in the visible universe on which $x$ def-depends are
  required with both their type and body.
\end{compact-itemize}

Let $S$ be a species containing the set of methods $\{ y_i : \tau_i = e_i \}$.
Let $x$ being one of the $y_i$, the minimal typing environment of
$x$ is defined as follows:

\vspace{-0.2cm}
{\scriptsize
$$
\emptyset \mintypenv x = \emptyset
\\
\hspace{1cm}
\inferrule
  { y \not\in\ \visuniv{x} \\ \{ y_i : \tau_i = e_i \} \mintypenv x = \Sigma }
  { \{ y : \tau = e\ ;\ y_i : \tau_i = e_i \} \mintypenv x = \Sigma }
$$
\vspace{-0.2cm}
$$
\inferrule
  { y \in \visuniv{x} \\ y \defdepbytrans{S} x \\
  \{ y_i : \tau_i = e_i \} \mintypenv x = \Sigma }
  { \{ y : \tau = e\ ;\ y_i : \tau_i = e_i \} \mintypenv x =
    \{ y : \tau = e \ ;\ \Sigma \} }
\\
\hspace{1cm}
\inferrule
  { y \in \visuniv{x} \\ y \not{\defdepbytrans{S}} x \\
  \{ y_i : \tau_i = e_i \} \mintypenv x = \Sigma }
  { \{ y : \tau = e\ ;\ y_i : \tau_i = e_i \} \mintypenv x =
    \{ y : \tau \ ;\ \Sigma \} }
$$
}

\vspace{-0.3cm}
Using the minimal typing environment it is possible to generate the method
generators for non-parametrised species only since collection parameters
are not taken into account. {\em A fortiori} it is not possible to generate
collection generators.

\subsubsection{On Methods from Parameters}
Following the same principle than in the previous section, we note
the direct dependencies of an expression $e$ in a species $S$ on a
parameter $C$ by ${\eddepsonparam S C e}$ and define it by a simple search
on the AST (looking for occurrences of the form \lstinline"C!x" for any $x$).
In the coming rule, ${\cal E}(S)$ stands for the parameters of the species
$S$ and ${\mbody S x}$ returns the body of the method $x$ in the species
$S$ (i.e. an expression for a {\tt let}-definition and a proof for a theorem).

\medskip
The challenge is to find the minimal set of parameters methods required to
typecheck a method. We now present the five first rules driving the calculus
since they do not have any order of application. A last one will be exposed
after.

\vspace{-0.2cm}
{\scriptsize
$${\bdepsonparam S C x} = {\eddepsonparam S C {\mbody S x}} \hspace{1.5cm} {\tdepsonparam S C x} = {\eddepsonparam S C {\mtype S x}}$$
$${\dddepsonparam S C x} = {\eddepsonparam S C {\mbody S z}}
  \ \ \ \ \forall z \mbox{ such as } z \defdepbytrans{S} x \hspace{0.9cm} {\udepsonparam S C x} = {\eddepsonparam S C {\mtype S z}}
  \ \ \ \ \forall z \in\ \visuniv{x}$$
$$
\inferrule*
  {
  {\cal E}(S) = (\ldots, C_p \kw{ is } ^{{\bf ...}}, \ldots,
     C_{p'} \kw{ is } S'(\ldots, C_p, \ldots)) \\
  {\cal E}(S') = (\ldots, C'_k\ \kw{ is }\  I'_k,\ldots) \\\\
  z \in {\tdepsonparam S {C_{p'}} x}\ \vee\ z \in {\bdepsonparam S {C_{p'}} x} \\
  (y : \tau_y) \in {\tdepsonparam {S'} {C'_k} z}
  }
  {
  (y : \tau_y[C'_k \mapsfrom C_p]) \in {\pdepsonparam S {C_p} x}
  }
$$
}

\vspace{-0.4cm}
The rule $\textsc{[Body]}$ (resp. $\textsc{[Type]}$) takes into account
dependencies on a method explicitly stated in the body (resp. type) of a
definition.

The rules $\textsc{[Def]}$ and $\textsc{[Univ]}$ serve to collect dependencies
on a parameter induced by the dependencies a method has inside its
hosting species. Note that methods $z$ introduced by the rule \textsc{[Def]}
are obviously included  in those introduced by \textsc{[Univ]}. In effect,
the visible universe is wider than only transitive def-dependencies and their
related decl-dependencies: if there is no def-dependencies then the relation
$\defdepbytrans{S}$ will be empty although decl-dependencies may lead to a
non-empty visible universe. The rule \textsc{[Def]} allows to inspect bodies.
The rule \textsc{[Univ]} allows to inspect types. Hence, any $z$ introduced by
\textsc{[Def]} will also has its type inspected by \textsc{[Univ]}.

Finally, the rule \textsc{[Prm]} applies to take into account dependencies of
a method on a previous parameter $C_p$ used as argument to build the current
parameter $C_{p'}$. It differs from the previous rules, since the result of the
calculus is not only a set of names: types are explicit. This is because the
type of the methods of this set differs from the one computed
during the typechecking of the species used as parameter.

If $C$ is an entity parameter, we set:
${\cal D}o{\cal P}_{\textsc{[xxx]}} = \{ C \}$, i.e. that the identifier of the
parameter is considered as it's only method.

\medskip
None of these rules took into account decl-dependencies that methods of
parameters have inside their own species and that are visible through
types. The following example show that using \lstinline"P!th0" to prove
{\tt th1} which only deals with \lstinline"P!f"  however needs to have
\lstinline"P!g" in the context.

\noindent
\begin{minipage}{7.2cm}
{\tiny
\begin{lstlisting}[language=MyCoq,mathescape=true,frame=none]
species A =
  signature f : Self -> int ;
  signature g : Self -> int ;
  property th0: all x : Self, f (x) = 0 /\ g (x) = 1 ;
end ;;
\end{lstlisting}
}
\end{minipage}\hskip2cm
\begin{minipage}{8cm}
{\tiny
\begin{lstlisting}[frame=none]
species B (P is A) =
  theorem th1 : all x : P, P!f (x) = 0
  proof = by property P!th0 ;
end ;;
\end{lstlisting}}
\end{minipage}

Note that because of the encapsulation process applied to collection parameters,
def-dependencies are never possible and only types are visible. Hence the
following rule serves to complete an initial set of dependencies ${\cal D}$.

\vspace{-0.2cm}
{\scriptsize
$$
\inferrule* [Right=Close]
  {
  {\cal E}(S) = (\ldots, C_p \kw{ is } I_p, \ldots) \\
  z \in {\cal D} (S, C_p)[x] \\
  (y : \tau_y) \in {\decldeps {\mtype {I_p} z}}_{I_p}
  }
  {
  (y : \tau_y[{\tt Self} \mapsfrom C_p]) \in {\cal D}^+ ({\cal D}, S, C_p)[x]
  }
$$
}

\vspace{-0.4cm}
Note that new dependencies brought by this rule cannot themselves require
applying this rule again to make them typecheckable. In effect, only
logical methods can introduce new dependencies and they only can depend on
computational methods whose types are ``ML-like'' ones, hence cannot introduce
methods names.

\subsection{Code generation}\label{ref-code-reuse}
Providing detailed algorithms implementing code generation is out of
the scope of this paper. Instead, we illustrate their expected
behaviour by showing the output obtained by compiling our above
example.

Code generation starts after resolution of inheritance and
late-binding, typing and dependency analyses. Note that issuing very
similar target codes should ease assessment. Thus the
code generation phase should be common to the two targets until
concrete syntaxes  are produced. Moreover a good sharing of
code alleviates the assessment task and eases control of code size and
reuse.  Therefore we try to maximize sharing.

As inheritance and late-binding must be resolved at compile-time to
ensure proof validity, it would be possible to generate code  for
collections only, but this prevents any sharing as shown below.
Code generation of
{\tt In\_5\_10} and {\tt In\_1\_8} use the last definition of
methods of {\tt IsIn} but they do not share their bodies. Some
possible sharing of the methods of the parameter
{\tt IntC} are also lost. Not only code size
increases but assessment takes longer since any collection should
be checked {\it in extenso}. Thus, code has to be generated for all
species. 

\smallskip

{\bf Method Generators}
In the following example, the species {\tt IsInE} redefines  {\tt filter}.
{\tiny
\begin{lstlisting}[frame=none]
species IsInE (X is OrdData, low in X, high in X) =
  inherit IsIn (X, low, high) ;
  let filter  = ...
end ;;
collection ExtIn_3_8 = implement IsInE (...) ; end ;;
\end{lstlisting}} 
{\tt lowMin} def-depends on {\tt filter}.  Its body, that is, the proof
of {\tt lowMin} in {\tt IsIn}, can be shared by all species inheriting
{\tt IsIn}, if they do not redefine {\tt filter}. But it cannot be shared
with {\tt IsInE} as the proof must be re-done. 
Thus,  redefinitions cancel sharing of def-dependencies.

Assume that a method {\tt IsIn!m} decl-depends on {\tt filter}. Then
all species inheriting {\tt IsIn!m} and not redefining {\tt m} can
share its body, up to the calls of {\tt filter}. Similarly the
collections {\tt In\_5\_10}, {\tt In\_1\_8} and {\tt ExtIn\_1\_8} can share
the methods of {\tt IntC}, up to the calls of methods on which {\tt IntC}
{\em decl-depends}. Thus, the body of a definition {\tt m} (function or proof)
of a species {\tt S} can be shared along inheritance, up to
the calls to methods of {\tt S} and of its parameters upon which {\tt
  m} decl-depends (thanks to the absence of cycles in
decl-dependencies). The sharing is done by abstracting, in {\tt m}'s
body, the names of these decl-dependencies, using the
standard technique of  $\lambda$-lifting\cite{Johnsson85llifting}.
The result of this  $\lambda$-lifting  is called  the {\em
  method generator} of {\tt m}.
 To obtain the final code of {\tt m}, this
generator  will have to be applied to the most recent values of its
decl-dependencies, provided by the species (or its descendants)  ({\em
  spec-arguments}) or by the effective parameters ({\em param-arguments}).

Moreover, if the logical target language (like \coq) requires explicit
polymorphism, representations of \lstinline"Self" and of parameters on
which a {\em method} decl-depends are also $\lambda$-lifted (leading
to parameters of type {\tt Set} in \coq\ and named with {\tt \_T} in
the example below).  As the methods and representation of a species
can depend on representations and methods of collection parameters,
$\lambda$-liftings of decl-dependencies upon parameters must be the
outermost abstractions.

The generated codes for species are grouped into modules of the
target languages, to enforce modularity and to benefit of a
convenient namespace mechanism. Now  either these modules contain all
the inherited and non-redefined method generators of the species (but
this leads to code duplication) or a method generator is created once, when a
method is (re)defined and appears only in the species that defines
it. We use the latter solution.

\smallskip

{\bf Example continued.}
The example of section \ref{ref-example} is pursued, using a \coq-like
syntax. \ocaml\ files are just obtained by removing some types,
properties and proofs and are not listed here. In method generators,
any occurrence of the name
of a decl-dependency {\tt m} is replaced by \lstinline"abst_m" and
abstracted.  If by adding a definition, this decl-dependency is turned
into a def-dependency, then \lstinline"abst_m" is bound to the
definition of {\tt m} by the construct \lstinline"abst_m :=" {\it Coq term}. 

\smallskip
\noindent {\em Generating code for species {\tt Data}}:
\label{code-data}
The defined method {\tt id} has no dependencies, hence trivially leads
to a simple definition.

{\tiny
\begin{lstlisting}[language=MyCoq,frame=none]
Module Data.
  Definition id : basics.string__t := "default".
End Data.
\end{lstlisting}}

\noindent {\em Generating code for species {\tt OrdData}}:
\label{code-comparable} 
The body of {\tt gt} decl-depends on {\tt eq}, {\tt lt} and
the representation as its inferred type is
${\tt Self} \rightarrow {\tt Self} \rightarrow {\tt bool}$. Note that
this body remains close to the source one.
Inherited and only declared methods induce no code. 

{\tiny
\begin{lstlisting}[language=MyCoq,frame=none]
Module OrdData.
  Definition gt (abst_T : Set) (abst_eq : abst_T -> abst_T -> basics.bool__t)
    (abst_lt : abst_T -> abst_T -> basics.bool__t) (x : abst_T) (y : abst_T) : basics.bool__t :=
       basics.not (abst_lt x y) && basics.not (abst_eq x y)
End OrdData.
\end{lstlisting}}

\noindent{\em Generating code for species {\tt TheInt}}:
\label{code-nativeint}
 This species redefines {\tt id}
and defines {\tt eq}, {\tt from\-Int} and {\tt lt}.  {\tt id} has
no dependencies whereas {\tt eq}, {\tt from\_int} and {\tt lt} only
have a def-dependency on the representation, whose value (the built-in
type {\tt int}) is bound
(\lstinline":=" construct) to {\tt abst\_T} in the corresponding
method generators.

The proof of {\tt ltNotGt} decl-depends on the representation. It
def-depends on {\tt gt} because it unfolds its definition. Note that {\tt
  lt} and {\tt eq} do not appear in this proof : unfolding of {\tt gt}
does not unfold them recursively. But typing {\tt ltNotGt} requires typing
{\tt gt} and thus typing {\tt  lt} and {\tt eq}. Hence, {\tt
  ltNotGt} has decl-dependencies on  {\tt  lt} and {\tt eq}, coming
from the def-dependency on {\tt gt}. So, {\tt  lt} and {\tt eq} must
be $\lambda$-lifted in {\tt ltNotGt},  to build the value
of {\tt abst\_gt} by applying the method generator {\tt gt} found in
the module {\tt OrdData} to its three arguments. Note that only the
types of {\tt  lt} and {\tt eq} are used by  {\tt ltNotGt}: these
methods can be redefined without impacting this theorem.

{\tiny
\begin{lstlisting}[language=MyCoq,frame=none,mathescape=true]
Module TheInt.
  Definition id : basics.string__t := "native int".
  Definition eq (abst_T := basics.int__t) (x : abst_T) (y : abst_T) : basics.bool__t := (basics._equal_0x x y).
  Definition fromInt (abst_T := basics.int__t) (x : basics.int__t): abst_T := x.
  Definition lt (abst_T := basics.int__t) (x : abst_T) (y : abst_T) : basics.bool__t := (basics._lt_0x x y).
  Theorem ltNotGt (abst_T : Set) (abst_eq : abst_T -> abst_T -> basics.bool__t)
    (abst_lt : abst_T -> abst_T -> basics.bool__t) (abst_gt := OrdData.gt abst_T abst_eq abst_lt):
    forall x  y : abst_T, Is_true ((abst_lt x y)) -> ~Is_true ((abst_gt x y)).
  apply $"Large\ Coq\ term\ generated\ by\ Zenon"$ ;
End TheInt.
\end{lstlisting}}

As illustrated by this example, the dependency calculus cannot be
reduced to a simple ``grep''. For any method, the analysis must
compute the sets of methods of the species and of the parameters which
are needed to elaborate the type and the value of its logical code (in
\coq, they are called visible universe and minimal
typing environment). This is the price to pay for having no errors in
target codes. 

\smallskip

{\bf Collection generators} 
Code generation for collections must create computational
runnable code and checkable logical code. Suppose that a collection
{\tt C} is built by encapsulating a complete species {\tt S}. The code of a
method {\tt S!m} is obtained by applying its method generator, say
{\tt G\_m}, to its effective arguments. The right version of this
method generator {\tt G\_m} comes from the last definition of {\tt m}
in the inheritance path ending on {\tt S}. The effective
spec-arguments of {\tt G\_m} can be retrieved from the species having
created {\tt G\_m} and from the instantiations of formal parameters
done during inheritance. These applications can be safely done as the
species {\tt S} is complete and as the dependency analysis provides an
ordering on all methods of a species, such that the $n^{th}$ method
depends only on the $(n-1)$ first ones.  The effective param-arguments
of {\tt G\_m} come from the application of the species {\tt S} to
effective collection parameters.

Thus a simple solution to generate collection code is to do it method
by method, by applying  method generators to their effective
arguments. These applications are computed at runtime in
the target languages. This solution allows us to generate only
the needed applications for a given method. But a possibility of
sharing is lost, when collections are issued from the same
parameterized species, applied to different effective collection
parameters (case of {\tt In\_5\_10} and {\tt In\_1\_8}). Then the
applications of the
method generators to the spec-arguments can be shared between all
these collections. Regarding memory use, the gain is small. But
regarding assessment processes, such an intermediate
step of applications represents a large gain as it avoids
having to
review several copies of the same code, applied to different
param-arguments. We retain this solution.
To ease code review, the applications to the spec-arguments are
grouped into a record (we assume that target languages offer records)
called {\em collection generator} while the $\lambda$-liftings of all
parameters decl-dependencies are moved outside the record body.  Thus
the material coming from the species is found in the body while the effective
parameters contribution appears in the  applications of
the body. 

It is possible to go further by replacing the bunch of
$\lambda$-liftings of parameter de\-pen\-den\-cies with a unique
$\lambda$-lifting
abstracting the collection parameter. Then the target modules should
be first-class values of the target languages with a certain kind of
subtyping as interfaces inclusion provides a simple but useful
subtyping. Even if our target languages have such features, it seems
better to leave the code generation model  open to a wide range of potential targets.
 
\smallskip

{\bf Example ended}
We continue the example of \ref{ref-example} by completing the module
{\tt TheInt} with the collection generator and its record type. 

\noindent
\begin{minipage}{7.4cm}
{\tiny
\begin{lstlisting}[language=MyCoq,frame=none]
Record me_as_species :=
  mk_record {
  rf_T : Set ;
  rf_id : basics.string__t ;
  rf_eq : rf_T -> rf_T -> basics.bool__t ;
  rf_fromInt : basics.int__t -> rf_T ;
  rf_lt : rf_T -> rf_T -> basics.bool__t ;
  rf_gt : rf_T -> rf_T -> basics.bool__t ;
  rf_ltNotGt : forall x  y : rf_T, Is_true (rf_lt x y) ->
    ~Is_true (rf_gt x y)
  }.
\end{lstlisting}}
\end{minipage}\hskip0.5cm
\begin{minipage}{8cm}
{\tiny
\begin{lstlisting}[language=MyCoq,frame=none]
Definition collection_create :=
  let local_rep := basics.int__t in
  let local_id := id in
  let local_eq := eq in
  let local_fromInt := fromInt in
  let local_lt := lt in
  let local_gt := OrdData.gt local_rep local_eq local_lt in
  let local_ltNotGt := ltNotGt local_rep local_eq local_lt in
  mk_record
    local_rep local_id local_eq local_fromInt local_lt
    local_gt local_ltNotGt.
\end{lstlisting}}
\end{minipage}

The type of each method of the species is recorded into a record field
labeled {\tt rf\_}, its value {\tt local\_} (obtained by a future
application of the collection generator to all the param-arguments)
has no more $\lambda$-lift.  Here, as this collection has no
parameter, there is no $\lambda$-lifting on the record itself. The value of {\tt local\_gt}
for example is obtained by applying the method generator coming from
{\tt OrdData} to its spec-arguments, whose values have already been
generated, thanks to the absence of cycles in dependencies. The
function {\tt mk\_record} builds the record out of these values.

\noindent {\em Generating code for species {\tt IsIn}.}
\label{code-thresholder}
The types of the fields are translations of the types of the methods
of the collection.  The dependency calculus  shows that the record
depends on the carrier of the parameter {\tt V}, on
the two value parameters {\tt minv} and {\tt maxv} and on the
method \lstinline"V!gt". These decl-dependencies are
$\lambda$-lifted.  For example, the type of {\tt rf\_lowMin}
is the translation of the property {\tt lowMin}, which 
decl-depends on {\tt V!gt}. The  abstraction on  {\tt V!gt} needed for
{\tt lowMin} is also done on the other fields. 

{\tiny
\begin{lstlisting}[language=MyCoq,frame=none]
Module IsIn.
  Record me_as_species (V_T : Set) (_p_minv_minv : V_T) (_p_maxv_maxv : V_T)
      (_p_V_gt : V_T -> V_T -> basics.bool__t) : Type :=
    mk_record {
    rf_T : Set ;
    rf_filter : V_T -> rf_T ;
    rf_getStatus : rf_T -> statut_t__t ;
    rf_getValue : rf_T -> V_T ;
    rf_lowMin :
      forall x : V_T, Is_true (basics._equal_ _ (rf_getStatus (rf_filter x)) Too_low) -> ~Is_true (_p_V_gt x _p_minv_minv)
    }.
\end{lstlisting}}

Next,  methods generators are created for methods defined in {\tt IsIn}.

{\tiny
\begin{lstlisting}[language=MyCoq,frame=none,mathescape=true]
  Definition getValue (_p_V_T : Set) (abst_T := (_p_V_T * statut_t__t)) (x : abst_T) : _p_V_T := basics.fst _ _ x.
  Definition getStatus (_p_V_T : Set) (abst_T := (_p_V_T * statut_t__t)) (x : abst_T) : statut_t__t := basics.snd _ _ x.
  Definition filter (_p_V_T : Set) (_p_V_lt : _p_V_T -> _p_V_T -> basics.bool__t)
    (_p_V_gt : _p_V_T -> _p_V_T -> basics.bool__t) (_p_minv_minv : _p_V_T) (_p_maxv_maxv : _p_V_T)
    (abst_T := (_p_V_T * statut_t__t)) (x : _p_V_T) : abst_T :=
    (if (_p_V_lt x _p_minv_minv) then (_p_minv_minv, Too_low)
      else (if (_p_V_gt x _p_maxv_maxv) then (_p_maxv_maxv, Too_high) else (x, In_range))).
  Theorem lowMin  (_p_V_T : Set) (_p_V_lt : _p_V_T -> _p_V_T -> basics.bool__t)
    (_p_V_gt : _p_V_T -> _p_V_T -> basics.bool__t)
    (_p_V_ltNotGt : forall x  y : _p_V_T, Is_true ((_p_V_lt x y)) -> ~Is_true ((_p_V_gt x y)))
    (_p_minv_minv : _p_V_T) (_p_maxv_maxv : _p_V_T) (abst_T := (_p_V_T * statut_t__t))
    (abst_filter := filter _p_V_T _p_V_lt _p_V_gt _p_minv_minv _p_maxv_maxv) (abst_getStatus := getStatus _p_V_T):
    forall x : _p_V_T,
      Is_true ((basics._equal_ _ (abst_getStatus (abst_filter x)) Too_low)) -> ~Is_true ((_p_V_gt x _p_minv_minv)).
  apply $"Large\ Coq\ term\ generated\ by\ Zenon"$ ;
\end{lstlisting}} 

Methods have no decl-dependencies on methods of {\tt
IsIn}, except {\tt lowMin} which has a def-de\-pen\-den\-cy on
{\tt filter}. The other decl-dependencies are on the representation of
{\tt IsIn} (and through it, on the one of {\tt V}), on {\tt V}'s methods and
on values {\tt minv} and {\tt maxv}.
The def-dependency of {\tt lowMin} leads to the binding
(\lstinline":=") of {\tt abst\_filter} to the application of the method
generator {\tt filter} to all its arguments, represented by abstracted
variables in the context.

The body of the collection generator {\tt IsIn!collection\_create}
are the applications of the method generators to their param-arguments
(no spec-arguments here). Then these param-arguments are 
$\lambda$-lifted. 

{\tiny
\begin{lstlisting}[language=MyCoq,frame=none]
 Definition collection_create (_p_V_T : Set) _p_minv_minv _p_maxv_maxv _p_V_lt _p_V_gt _p_V_ltNotGt :=
   let local_rep := (_p_V_T * statut_t__t) in
   let local_filter := filter _p_V_T _p_V_lt _p_V_gt _p_minv_minv _p_maxv_maxv in
   let local_getStatus := getStatus _p_V_T in
   let local_getValue := getValue _p_V_T in
   let local_lowMin := lowMin _p_V_T _p_V_lt _p_V_gt _p_V_ltNotGt _p_minv_minv _p_maxv_maxv in
   mk_record
     (_p_V_T : Set) _p_minv_minv _p_maxv_maxv _p_V_gt local_rep local_filter local_getStatus local_getValue local_lowMin.
End IsIn.
\end{lstlisting}}

\noindent {\em Generating code for collection {\tt IntC}}:
The module generated from {\tt IntC} contains all the definitions
obtained by just extracting the fields of the collection generator
{\tt TheInt.collection\_create} as there are no 
parameters.\label{code-intcoll}

{\tiny
\begin{lstlisting}[language=MyCoq,frame=none]
Module IntC.
  Let effective_collection := TheInt.collection_create.
  Definition me_as_carrier := basics.int__t.
  Definition id := effective_collection.(TheInt.rf_id).
  Definition eq := effective_collection.(TheInt.rf_eq).
  Definition fromInt := effective_collection.(TheInt.rf_fromInt).
  Definition lt := effective_collection.(TheInt.rf_lt).
  Definition gt := effective_collection.(TheInt.rf_gt).
  Definition ltNotGt := effective_collection.(TheInt.rf_ltNotGt).
End IntC.
\end{lstlisting}}

\noindent {\em Generating code for collection {\tt In\_5\_10}}:
\label{code-thresh510}
Here, {\tt IsIn.collection\_create} is applied to effective arguments
found in the module {\tt IntC} and definitions are extracted as above.
The four underscores are  just arguments  inferred by \coq, which denote
the four parameters of the record.

{\tiny
\begin{lstlisting}[language=MyCoq,frame=none]
Module In_5_10.
  Let effective_collection :=
    IsIn.collection_create IntC.me_as_carrier (IntC.fromInt 5) (IntC.fromInt 10) IntC.lt IntC.gt IntC.ltNotGt.
  Definition filter := effective_collection.(IsIn.rf_filter _ _ _ _ ).
  Definition getStatus := effective_collection.(IsIn.rf_getStatus _ _ _ _).
  Definition getValue := effective_collection.(IsIn.rf_getValue _ _ _ _).
  Definition lowMin := effective_collection.(IsIn.rf_lowMin _ _ _ _).
End In_5_10.
\end{lstlisting}}

\subsection{Summarizing Dependencies Usage in $\lambda$-lifting}
As previously introduced, dependencies are subject to be $\lambda$-lifted to
define record types, method generators and collection generators. In a
symmetrical fashion, they determine the effective methods to provide to these
generators, which can only be achieved taking care of the instantiations of
formal collection and entity parameters by effective arguments along the
inheritance. The detailed mechanism of this is out of the scope of the present
paper. Instead, we summarize here the material to $\lambda$-lift for each
category of code generated element, in order of apparition for a species:

\begin{compact-itemize}
\item {\bf Parameters carriers} For record type and collection generator: all
  those of the parameters. For method generators: per method, only those of
  the used parameters.
\item {\bf Parameters methods}
  \begin{compact-itemize}
    \item For record type: ``union'' of all the dependencies of all the
      methods got by \textsc{[Close]} (\textsc{[Type]}).
    \item For method generators: the dependencies of the related method
      obtained by rules:
      \textsc{[Body]} + \textsc{[Type]} + \textsc{[Close]} (\textsc{[Def]} +
      \textsc{[Univ]} + \textsc{[Prm]}).
    \item For collection generator: ``union'' of methods dependencies
      abstracted in each of the method generators.
  \end{compact-itemize}
\item {\bf Methods of {\tt Self}} Only needed for method generators: those
  belonging to the minimal typing environment that are only declared.
\end{compact-itemize}

Note that because of relative dependencies between methods of parameters inside
their own species, $\lambda$-lifts of methods must be ordered for a same
parameter to ensure they only refer to previously $\lambda$-lifted elements.
Moreover, parameters are processed in order of apparition in the species: this
way, all the methods of a same parameter are $\lambda$-lifted in sequence.

\section{Conclusion}
Building \focal, a lot of questions,
choice points, etc. arose from the will to avoid dissociating the
computational and logical aspects in a formal development while
keeping the FIDE palatable. The mix of inheritance, late binding,
encapsulation, inductive data types and unfolding of definitions in
proofs creates complex dependencies, which have to be analyzed to
ensure development correctness. This analysis gives the basis of the
compilation process  through the notions of method and collection
generators, that we introduced to allow code sharing. The code
generation model producing computable and logical target codes is
outlined through an example. The formal computation of dependencies was
presented with an short summary of their usage in the $\lambda$-lifting
process.

Several other FIDEs mentioned in the introduction have followed other
choices. It should be very interesting to compare their compilation
models with ours, particularly on the method used to establish
correspondence between runnable and logical code and on their
semi-automation of proofs.

\zenon\ greatly contributes to \focal\ since it brings proof automation. This point is especially crucial to keep
proofs simpler.
 We plan to extend it to handle recursion termination,
arithmetic, temporal properties and, as it targets other provers
than \coq, to experiment with other target type theories.  

A huge amount of work remains to do in order to enhance \focal.
But, with all its weaknesses, it has already proved to be efficient in
non-trivial developments. While safety domains already have a large number of
good tools, security domains are much less well endowed, and the recent
interest in combining safety and security requirements will increase demand for
such tools.

\vspace{-0.3cm}
\bibliographystyle{eptcs}
\bibliography{bibli}
\end{document}